\documentstyle[hep93]{article}
\begin{document}
\title{Soft Terms from Strings
\thanks{Based on talks given at the 29th International Symposium on the
Theory of Elementary Particles, Buckow (Berlin), August 1995; 
5th Hellenic School 
and Workshops on Elementary Particle Physics, Corfu, September 1995.}}
\author{C. Mu\~noz\thanks{Research supported in part by: the CICYT, 
                                   under contract AEN93-0673; the European 
                                   Union, under contracts CHRX-CT93-0132 and 
                                   SC1-CT92-0792.}\\
        {\em Departamento de F\'{\i}sica Te\'orica C-XI, 
                           Universidad Aut\'onoma de Madrid}\\
        {\em Cantoblanco, 28049 Madrid, Spain}\\
        {\em cmunoz@ccuam3.sdi.uam.es}
       }
\abstract{
We study the structure of the soft SUSY-breaking terms obtained
from 4-D Strings
under the assumption of dilaton/moduli dominance in the process of
SUSY breaking.
We first 
analyze in detail the
dilaton-dominated limit because of its finiteness properties and 
phenomenological predictivity, and second, we consider the new
features appearing when several moduli fields contribute to SUSY breaking. 
In particular, we discuss in detail the case of symmetric Abelian orbifolds.
Although some qualitative
features indeed change in
the multimoduli case with respect to the dilaton dominance one, the
most natural mass relations at low-energy, 
$m_l < m_q \simeq M_g$, are still similar. Only in some very specific limits
these relations might be reversed.
We also study the presence of tachyons pointing out that
their possible existence may be, in some cases, 
an interesting advantage in order to break extra gauge symmetries.
Finally, we compute explicitly the $\mu$ and $B$ parameters in the
context of the mechanism
for generating a ``$\mu $-term'' by the K\"ahler potential, 
as  naturally implemented in orbifolds. It
leads to the prediction $|tg\beta |=1$ at the String scale,
independently of the Goldstino direction.
It is worth noticing that in this squeme the dilaton-dominated case,
where there is no free parameters, is excluded since it is not consistent
with the measured value of the top-quark mass.
In this connection, low-energy charge and color breaking minima are also
discussed.
}
\maketitle
\thispagestyle{empty}

\leftline{}
\leftline{}
\leftline{}
\leftline{}
\leftline{FTUAM 96/4}
\leftline{hep-ph/9501???}
\leftline{January 1996}
\leftline{}

\vskip-17.cm
\rightline{}
\rightline{ FTUAM 96/4}
\rightline{hep-ph/9501???}
\vskip2in

\newpage
\setcounter{page}{1}

\section{INTRODUCTION AND SUMMARY}

In the last four years there has been much activity in 
trying to obtain information
about the structure
of soft Supersymmetry (SUSY)-breaking
terms in effective $N=1$ theories coming from
four-dimensional (4-D) Strings \cite{review}. 
The basic idea is to identify some $N=1$ chiral
fields
whose
auxiliary components could break SUSY by acquiring a vacuum expectation
value (VEV).
Natural
candidates in 4-D Strings are 1) the complex dilaton field
$S={{1}\over {g^2}}
+ia$ which is present in any 4-D String and 2) the moduli fields
$T^i, U^i$ which parametrize the size and shape of the compactified variety
in models obtained by compactification of a ten-dimensional heterotic String.
It is not totally unreasonable to think that some of these fields may play an
important role
in SUSY breaking. To start with, if String models are to make any sense, these
fields
should be strongly affected by non-perturbative phenomena. They are massless in
perturbation
theory and non-perturbative effects should give them a mass to avoid deviations
from the equivalence principle and other phenomenological problems.
Secondly, these fields are generically present in large classes of
4-D 
models (the dilaton in all of them). Finally, the couplings of these fields to
charged matter
are suppressed by powers of the Planck mass, which makes them natural
candidates
to
constitute the SUSY-breaking ``hidden sector'' which is assumed to be present
in phenomenological models of low-energy SUSY.

The important point in this assumption of locating the seed of
SUSY-breaking
in the dilaton/moduli sectors, is that it leads to some interesting
relationships among different soft terms [2-7] 
which could perhaps be experimentally
tested. This general analysis was applied in particular to the
gaugino condensation scenario 
in ref.\cite{gaugino}, whereas in 
refs.[3-7] 
no special assumption was made about the possible origin of SUSY breaking.

In ref.\cite{BIM} a systematic discussion of
the structure of soft terms which may be obtained under the assumption of
dilaton/moduli-dominated SUSY breaking in some classes of 4-D 
Strings was presented,
with particular emphasis on the case of Abelian $(0,2)$ orbifold 
models \cite{orbifolds}.
It was mostly considered a situation in which only the dilaton $S$ and
an ``overall modulus $T$'' field contribute to SUSY breaking.
In fact, actual 4-D Strings like orbifolds contain several $T_i$
and $U_i$ moduli.
Generic $(0,2)$ orbifold models contain three $T_i$ moduli
fields (only $Z_3$ has 9 and $Z_4$, $Z_6'$ have 5) and  a maximum of three
(``complex structure'') $U_i$ fields. The use of an overall modulus $T$ is
equivalent to the assumption that the three $T_i$ fields of generic orbifold
models contribute exactly the same to SUSY breaking and the rest do not 
contribute. In the absence
of further dynamical information it is reasonable to expect
similar contributions from {\it all} the 
moduli although not necessarily exactly
the same. 
Thus it is natural to ask what changes if one
relaxes the overall modulus hypothesis and works with the multimoduli 
case \cite{BIMS}.
This is one of the purposes of the present talk.

The second one is to analyze in more detail the dilaton-dominated limit, where
only the dilaton field contributes to SUSY breaking \cite{KL,BIM}. 
This is a very interesting
possibility not only due to phenomenological reasons, as universality of
the soft terms, but also to theoretical arguments.
In this connection it has recently been realized \cite{tonto,I,jeta} that 
the  boundary conditions $-A=M_{1/2}={\sqrt{3}}m$
of dilaton dominance coincide with some boundary
conditions considered by Jones, Mezincescu and Yau
in 1984 \cite{JMY} in a complete different context. 
They found that those same boundary conditions
mantain the (two-loop) finiteness properties of
certain $N=1$ SUSY theories \cite{finitas}. 
This coincidence is in principle quite surprising since we did not bother
about the loop corrections when extracting these boundary conditions from the
dilaton-dominance assumption. Also, effective $N=1$ field theories from
Strings do not in general fulfil the finiteness requirements. It has also
been noticed \cite{I} that this coincidence could be related to an underlying
$N=4$ structure of the dilaton Lagrangian and that the dilaton-dominated
boundary conditions could appear as a fixed point of renormalization
group equations \cite{I,J}.
This could perhaps be an indication that at least
some of the possible soft terms obtained in
the present scheme could have a more general
relevance, not necessarily linked to a particular form of the tree-level
Lagrangian.

In section 2 we present an analysis of the effects of several moduli
on the results obtained for soft terms.
In the multimoduli case several parameters are needed to specify the
Goldstino direction in the dilaton/moduli space, in contrast with
the overall modulus case where the relevant information is contained
in just one angular parameter $\theta$. The presence of more free
parameters leads to some loss of predictivity for the soft terms.
This predictivity is recovered and increased in the case of dilaton
dominance  where the soft terms, eq.(\ref{dilaton}), 
are independent of the 4-D String
considered and fulfil the low-energy mass relations
given by eq.(\ref{dilaton2}).
Also we show that, even in the multimoduli case, in some schemes 
there are certain
sum-rules among soft terms, eq.(\ref{rulox}), which hold independently
of the Goldstino direction.
The presence of these sum rules  cause that,
{\it on average} the
{\it qualitative} results in the dilaton-dominated case 
still apply. Specifically, if one insists e.g. in
obtaining scalar masses heavier than gauginos
(something not possible 
in the dilaton-dominated scenario),
this is possible in the multimoduli case, but
the sum-rules often force some of the scalars
to get negative squared mass.
If we want to avoid this, we have to
stick to gaugino masses bigger than
(or of order) the scalar masses.
This would lead us back to the qualitative results
obtained in dilaton dominance.
Let us notice however that in very specific limits, which will be discussed
below, these results might
be modified.
In the case of
standard model 4-D Strings the 
tachyonic behaviour may be particularly problematic,
since charge and/or colour could be broken.
In the case of GUTs constructed from Strings,
it may just be the signal
of  GUT symmetry breaking.
However, even in this case one expects the same order of
magnitude results for observable scalar and gaugino masses and hence the most
natural mass relations
{\it at low-energy} are still similar to the dilaton dominance ones.

Another topic of interest is the $B$ parameter, the soft mass term
which is associated to a SUSY-mass term $\mu H_1H_2$ for the
pair of Higgses $H_{1,2}$ in the Minimal Supersymmetric Standard Model (MSSM).
Compared to the other soft
terms, the result for the $B$ parameter is more model dependent.
Indeed, it depends not only on the dilaton/moduli-dominance assumption but also on the particular mechanism which could
generate the associated ``$\mu$ term'' \cite{review}. 
An interesting possibility to
generate such a term is the one suggested in refs.\cite{GM,CM}
in which it was
pointed out that in the presence of certain bilinear terms in the
K\"ahler potential an effective $\mu$ term of order the gravitino
mass, $m_{3/2}$, is naturally
generated. Interestingly enough, such bilinear terms in the
K\"ahler potential do appear in String models and particularly in Abelian orbifolds. In section 3 we compute the $\mu $ and
$B$ parameters\footnote{The results for $B$ corresponding to
the possibility of generating a small
$\mu$ term from the superpotential \cite{CM}
can also be found, for the multimoduli case
under consideration, in ref.\cite{BIMS}. 
They are more model dependent.} 
as well as the soft scalar masses of the
charged fields which could play the role of Higgs particles in
such Abelian orbifold schemes. We find the interesting result that,
{\it independently of the Goldstino direction} in the
dilaton/moduli space, one gets the prediction $|tg\beta |=1$
at the String scale. 
On the other hand, if we consider the interesting Goldstino direction
where only the dilaton breaks SUSY, the whole soft terms and the 
$\mu$ parameter depend only on the gravitino mass. Imposing the
phenomenological requirement of correct electroweak breaking we arrive
to the remarkable result that the whole SUSY spectrum is completely 
determined with no free parameters.
Unfortunately, this direction is not consistent with the measured value
of the top-quark mass. In this connection, an interesting comment about
low-energy charge and color breaking minima in the dilaton-dominated limit
can be found at the end of the section.

A few comments before closing up this summary are in order. First of all we are
assuming here that the seed of SUSY breaking propagates through the
auxiliary fields of the dilaton $S$ and the moduli $T_i$, $U_i$ fields.
However attractive  this possibility might be, it is fair to say that
there is no compelling reason why indeed no other fields in the
theory could participate. Nevertheless the present scheme
has a certain predictivity due to the relative universality
of the couplings of the dilaton and moduli. Indeed, the
dilaton has universal and model-independent couplings which are
there independently of the 4-D String considered.
The moduli $T_i$, $U_i$ fields are less universal, their number and structure
depend on the type of compactification considered. However, there are
thousands of different $(0,2)$ models with different particle content
which share the same $T_i$, $U_i$ moduli structure. For example, the moduli
structure of a given $Z_N$ orbifold is the same for all the thousands
of $(0,2)$ models one can construct from it by doing different
embeddings and adding discrete Wilson lines. So, in this sense,
although not really universal, there are large classes of models with
identical $T_i$, $U_i$ couplings.
This is not the case of generic charged matter fields
whose number and couplings are completely out of control,
each individual model being in general completely different from any other.
Thus assuming dilaton/moduli dominance in the SUSY-breaking
process has at least the advantage of leading to specific
predictions for large classes of models whereas if charged
matter fields play an important role in SUSY breaking we
will be forced to a model by model analysis,
something which looks out of reach.

Another point to remark is that we will use the tree level forms
for both the gauge kinetic function and the K\"ahler potential.
One-loop corrections to these functions have been computed in 
refs.\cite{loop} and \cite{loop2} respectively
in some classes of 4-D Strings (orbifold models) and 
their effects on the soft terms have also
been studied \cite{BIM,japoneses,BC} and could be
included in the analysis without difficulty. In fact,
the effects of these one-loop corrections will in general
be negligible except for those corners of the Goldstino
directions in which the tree-level soft terms vanish.
However, as we will see below, this situation would be a sort
of fine-tuning.
More worrysome are the possible non-perturbative
String corrections to the K\"ahler and gauge kinetic functions.
We have made use in our orbifold models of the
known tree-level results for those functions.
If the non-perturbative String  corrections turn out to be important,
it would be impossible to make any prediction about
soft terms unless we know all the relevant non-perturbative
String dynamics, something which looks rather remote
(although perhaps not so remote as it looked one year ago!).

\section{SOFT TERMS}

\subsection{General structure: the multimoduli case}
\label{subsec:general}

We are going to consider $N=1$ SUSY 4-D Strings with
$m$ moduli $T_i$, $i=1,..,m$. Such notation refers to both $T$-type
and $U$-type (K\"ahler class and complex structure in the Calabi-Yau
language) fields.
In addition there will be charged matter fields $C_{\alpha }$ and the
complex dilaton field $S$.
In general we will be considering $(0,2)$ compactifications and thus the
charged
fields do not need to correspond to $27$s of $E_6$.

Before further specifying the class of theories that we are going to consider
a comment about the total number of moduli is in order.
We are used to think of large numbers of $T$ and $U$-like moduli
due to the fact that in $(2,2)$ ($E_6$) compactifications there is a
 one to one correspondence between moduli and charged
fields. However, in the case of $(0,2)$ models
with arbitrary gauge group (which is the case of
phenomenological interest) the number of moduli is drastically reduced.
 For example,
in the standard $(2,2)$ $Z_3$ orbifold there are 36 moduli $T_i$,
9 associated to the untwisted sector and 27 to the fixed points of the
orbifold.
In the thousands of $(0,2)$ $Z_3$ orbifolds one can construct by adding
different
gauge backgrounds or doing different gauge embeddings, only the
9 untwisted moduli remain in the spectrum. 
The same applies to models with $U$-fields. This is also the case
for compactifications using $(2,2)$ minimal superconformal models. Here all
singlets associated to twisted sectors are projected out when proceeding
to $(0,2)$ \cite{plesser}.
So, as these examples
show, in the case of $(0,2)$ compactifications
 the number of moduli is drastically reduced
to a few fields.
In the case of generic Abelian orbifolds one is in fact left with
only three T-type moduli $T_i$ ($i=1,2,3$), the only exceptions being
$Z_3$, $Z_4$ and $Z'_6$, where such number is 9, 5 and 5 respectively.
The number of $U$-type fields in these $(0,2)$ orbifolds oscillates
between $0$ and $3$, depending on the specific example.
Specifically, $(0,2)$ $Z_2\times Z_2$ orbifolds have 3 $U$ fields,
the orbifolds of type $Z_4,Z_6$,$Z_8,Z_2\times Z_4$,$Z_2\times Z_6$ and
$Z_{12}'$ have just one $U$ field and the rest have no untwisted $U$-fields.
Thus, apart from the three exceptions mentioned above,
this class of models has at most 6 moduli, three of $T$-type (always
present) and at most three of $U$-type. In the case of models obtained
from Calabi-Yau type of compactifications
a similar effect is expected and  only one $T$-field associated to the
overall modulus is guaranteed to exist in $(0,2)$ models.

We will consider effective $N=1$ supergravity (SUGRA)
K\"ahler potentials of the
type:
\begin{eqnarray}
& K(S,S^*,T_i,T_i^*,C_{\alpha},C_{\alpha}^*)\ = \
-\log(S+S^*)\ +\ {\hat K}(T_i,T_i^*)\
+\
{\tilde K}_{{\overline{\alpha }}{ \beta }}(T_i,T_i^*){C^*}^{\overline {\alpha}}
C^{\beta }\ + (Z_{{\alpha }{ \beta }}(T_i,T_i^*){C}^{\alpha}
C^{\beta }\ +\ h.c. \ ). &
\label{kahl}
\end{eqnarray}
The first piece is the usual term corresponding to the complex
dilaton $S$ which is present for any compactification whereas the
second is the K\"ahler potential of the moduli fields, where we recall
that we are denoting
the $T$- and $U$-type moduli collectively by $T_i$.
The greek indices label the matter fields and their
kinetic term functions are  given by
${\tilde K_{{\overline{\alpha }}{ \beta }}}$ and $Z_{{\alpha }{\beta }}$
to lowest order in the matter fields. The last piece is often forbidden
by gauge invariance in specific models although it may be relevant
in some cases as discussed in section 3.
The complete $N=1$ SUGRA Lagrangian is determined by
the K\"ahler potential $K({\phi }_M ,\phi^*_M)$, the superpotential
$W({\phi }_M)$ and the gauge kinetic functions
$f_a({\phi }_M)$, where $\phi_M$ generically denotes the chiral fields
$S,T_i,C_{\alpha }$. As is well known, $K$ and $W$ appear in the
Lagrangian only in the combination $G=K+\log|W|^2$. In particular,
the (F-part of the) scalar potential is given by
\begin{eqnarray}
& V(\phi _M, \phi ^*_M)\ =\
e^{G} \left( G_M{K}^{M{\bar N}} G_{\bar N}\ -\ 3\right) \ , &
\label{pot}
\end{eqnarray}
where $G_M \equiv \partial_M G \equiv \partial G/ \partial \phi_M$
and $K^{M{\bar N}}$ is the inverse of the K\"ahler metric
$K_{{\bar N }M}\equiv{\partial}_{\bar N}{\partial }_M K$.

The crucial assumption  now is  to locate the origin of SUSY breaking in the
dilaton/moduli sector. 
Then, plugging eq.(\ref{kahl}) into eq.(\ref{pot}), 
the bosonic soft SUSY-breaking terms can be computed. Applying the
standard SUGRA formulae \cite{Soni} to the most general case where the
moduli and matter metrics are not diagonal we obtain:
\begin{eqnarray}
& {m}'^2_{{\overline{\alpha }}{ \beta }} = 
m_{3/2}^2 {\tilde K_{{\overline{\alpha }}{ \beta }}}
- {\overline F}^{\overline{i}} ( \partial_{\overline{i}}\partial_j
{\tilde K_{{\overline{\alpha }}{ \beta }}}
-\partial_{\overline{i}} {\tilde K_{{\overline{\alpha }}{ \gamma}}}
{\tilde K^{{ \gamma} {\overline{\delta}} }}
\partial_j {\tilde K_{{\overline{\delta}}{ \beta}}}  ) F^j \ , &
\label{mmatrix}
\\
& A'_{\alpha\beta\gamma} = 
F^S K_S h_{\alpha\beta\gamma} + F^i \left[ {\hat K}_i h_{\alpha\beta\gamma}
+ \partial_i h_{\alpha\beta\gamma} - \left(
{\tilde K^{{ \delta} {\overline{\rho}} }}
\partial_i {\tilde K_{{\overline{\rho}}{ \alpha}}} h_{\delta\beta\gamma}
+(\alpha \leftrightarrow \beta)+(\alpha \leftrightarrow \gamma)\right)\right] 
\ , &
\label{mmatrix2}
\end{eqnarray}
where ${m}'^2_{ {\overline{\alpha }} { \beta } }$
and $A'_{\alpha\beta\gamma}$ are the soft mass matrix and the soft trilinear parameters respectively (corresponding to un-normalized charged
fields), $h_{\alpha \beta \gamma }$ is a (un-rescaled) renormalizable
Yukawa coupling involving three charged chiral fields and
$F^S=e^{G/2} K_{ {\bar{S}} S}^{-1} G_{\bar{S}}$, 
$F^i=e^{G/2} {\hat K}^{i {\overline j}} G_{\overline j}$ are the dilaton and
moduli auxiliary fields.
Notice that, after normalizing the fields to get
canonical kinetic terms, the first piece in
eq.(\ref{mmatrix}) will lead to universal diagonal soft
masses but the second piece will generically induce
off-diagonal contributions. Concerning the
$A$-parameters, notice that we have not
factored out the Yukawa couplings as usual, since
proportionality is not guaranteed.
Indeed, although the first term in
$A'_{\alpha\beta\gamma}$ is always proportional
in flavour space to the corresponding Yukawa
coupling, the same thing is not necessarily true
for the other terms.
In this section we are going to consider the case of diagonal metric
both for the moduli and the matter 
fields\footnote{An extensive 
analysis of the off-diagonal case in specific orbifold  constructions, including the calculation
of the soft terms and their effects on flavour changing neutral currents (FCNC),
can be found in ref.\cite{BIMS}.}.
Then ${\hat K}(T_i,T_i^*)$ will be a sum
of contributions (one for each $T_i$), whereas
${\tilde K_{{\overline{\alpha }}{ \beta }}}$ will be taken of the
diagonal form ${\tilde K_{{\overline{\alpha }}{ \beta }}}
\equiv \delta _{{\overline{\alpha }}{ \beta }} {\tilde K_{\alpha }}$.

Let us take the following parametrization
for the VEV's of the dilaton and moduli auxiliary
fields 
\begin{eqnarray}
 & G_{ {\bar{S}} S}^{1/2} F^S\ =\ \sqrt{3}m_{3/2}\sin\theta 
e^{-i\gamma _S}\ \ , &
\nonumber \\
&  G_{ {\bar{i}} i}^{1/2} F^i\ =\ \sqrt{3}m_{3/2}\cos\theta\ e^{-i\gamma _i}
\Theta _i \ \ , & 
\label{auxi}
\end{eqnarray}
where $\sum _i \Theta _i^2=1$ and $e^G=m^2_{3/2}$ is the gravitino
mass-squared.
The angle $\theta $ and the $\Theta _i$ just parametrize the
direction of the goldstino in the $S,T_i$ field space.
 We have also allowed for the possibility of
some complex phases $\gamma _S, \gamma _i$ which could be relevant
for the CP structure of the theory. This parametrization has the virtue that
when we plug it in the general form of the SUGRA scalar potential
eq.(\ref{pot}), its VEV (the cosmological constant) vanishes by
construction. Notice that such a phenomenological approach allows us
to `reabsorb' (or circumvent) our ignorance about the (nonperturbative)
$S$- and $T_i$- dependent part of the superpotential, which is
responsible for SUSY breaking.

It is now a straightforward
exercise
to compute the bosonic soft SUSY-breaking terms in this class of theories.
Plugging
eq.(\ref{auxi}) into eqs.(\ref{mmatrix},\ref{mmatrix2})
one finds the following results (we recall that we
are considering here a diagonal metric for the matter fields):
\begin{eqnarray}
 & m_{\alpha }^2 = \  m_{3/2}^2 \ \left[ 1\ -\ 3\cos^2\theta \
({\hat K}_{ {\overline i} i})^{-1/2} {\Theta }_i e^{i\gamma _i} 
(\log{\tilde K}_{\alpha })_{ {\overline i} j}
({\hat K}_{ {\overline j} j})^{-1/2} {\Theta }_j e^{-i\gamma _j} \ \right] \ ,
&
\nonumber \\
& A_{\alpha \beta \gamma } =
 \   -\sqrt{3} m_{3/2}\ \left[ e^{-i{\gamma }_S} \sin\theta 
- \ e^{-i{\gamma }_i} \cos\theta \  \Theta_i
({\hat K}_{ {\overline i} i})^{-1/2} \left({\hat K}_i 
- \sum_{\delta=\alpha,\beta,\gamma}
(\log {\tilde K}_{\delta })_i
+ (\log h_{\alpha \beta \gamma } )_i \ \right)
\  \right] \ . &
\label{soft}
\end{eqnarray}
The above scalar masses and trilinear scalar couplings (where we have factorized out the Yukawa coupling as usual) correspond
to charged fields which have already been canonically normalized.

Physical gaugino masses $M_a$ for the canonically normalized gaugino fields
are given in general by 
$M_a=F^M[log(Re f_a)]_M$.
Since the tree-level gauge kinetic function is given for any 4-D String by
$f_a=k_aS$, where $k_a$ is the Kac-Moody level of the gauge factor,
the result for tree-level gaugino masses is independent of the
moduli sector and is simply given by:
\begin{eqnarray}
& M\equiv M_a\ =\ m_{3/2}\sqrt{3} \sin\theta e^{-i\gamma _S} \ . &
\label{gaugin}
\end{eqnarray}

As we mentioned above, the parametrization of the auxiliary field VEV's
was chosen in such a way to guarantee the automatic vanishing of
the VEV of the scalar potential ($V_0=0$). If the value of $V_0$
is not assumed to be zero
the above formulae (\ref{auxi}-\ref{gaugin}) are modified in the following simple way.
One just has to replace $m_{3/2}\rightarrow Cm_{3/2}$,
where $|C|^2=1+V_0/3m_{3/2}^2$. In addition, the formula for $m_{\alpha }^2$
gets an additional contribution given by $2m_{3/2}^2(|C|^2-1)=2V_0/3$.

The soft term formulae above (\ref{soft}, \ref{gaugin}) 
are in general valid for any compactification
as long as we are considering diagonal metrics. In addition one is tacitally
assuming that the tree-level K\"ahler potential and $f_a$-functions
constitute a good aproximation.
The K\"ahler potentials
for the moduli are in general complicated functions.
Before going into specific classes of Superstring models, it is worth 
studying the interesting limit 
$\cos\theta =0$, corresponding to the case where the dilaton sector
is the source of all the SUSY breaking (see eq.(\ref{auxi})).

\subsection{The $\cos\theta =0$ (dilaton-dominated) limit}
\label{subsec:dilaton}

Since the dilaton couples in an universal manner to all particles, 
{\it this limit is quite model independent}. Using 
eqs.(\ref{soft},\ref{gaugin}) 
one finds the following simple expressions for the soft terms which are
independent of the 4-D String considered
\begin{eqnarray}
 & m_{\alpha } = \  m_{3/2} \ , &
\nonumber \\
 & M_a = 
 \ \pm \sqrt{3} m_{3/2}  \ , &
\nonumber \\
 & A_{\alpha \beta \gamma } =
 \   - M_a , &
\label{dilaton}
\end{eqnarray}
where, from the limits on the electric dipole moment of the neutron, we have
imposed $\gamma_S$ = $0$ mod $\pi$.

This dilaton-dominated scenario \cite{KL,BIM} 
is attractive for its simplicity and for
the natural explanation that it offers to the universality of the
soft terms. Actually, universality is a desirable property not
only to reduce the number of independent parameters in the MSSM, but also
for phenomenological reasons, particularly to avoid FCNC. 

Because of the simplicity of this scenario, the low-energy predictions 
are quite precise \cite{BLM,BIM,Vissani}. 
Since scalars are lighter than gauginos 
at the String scale, at low-energy ($\sim M_Z$), gluino, slepton and
(first and second generation) squark mass relations turn out to 
be\footnote{The phenomenology of SUSY breaking by the dilaton in the context
of a flipped $SU(5)$ model was also studied in ref.\cite{Nano}.}
\begin{eqnarray}
& M_g:m_Q:m_u:m_d:m_L:M_e  \simeq 1:0.94:0.92:0.92:0.32:0.24 \ . &
\label{dilaton2}
\end{eqnarray}
Although squarks and sleptons have the same soft mass, at low-energy the
former are much heavier than the latter because of the gluino contribution to
the renormalization of their masses.


\subsection{Orbifold compactifications}\label{subsec:orbifold}

To illustrate some general features of the multimoduli case
 we will concentrate here on the case of generic $(0,2)$ symmetric
Abelian orbifolds. As we mentioned above, this class of models
contains three $T$-type  moduli and (at most) three $U$-type moduli.
We will denote them collectively by $T_i$, where e.g. $T_i=U_{i-3}$; $i=4,5,6$.
For this  class of models the K\"ahler potential has the 
form \cite{potential}
\begin{eqnarray}
& K(\phi,\phi^*)\ =\ -\log(S+S^*)\ -\ \sum _i \log(T_i+T_i^*)\ 
+ \sum _{\alpha } |C_{\alpha }|^2 \Pi_i(T_i+T_i^*)^{n_{\alpha }^i} \ . &
\label{orbi}
\end{eqnarray}
Here $n_{\alpha }^i$ are fractional numbers usually called ``modular weights"
of the matter fields $C_{\alpha }$. For each given Abelian orbifold,
independently of the gauge group or particle content, the possible
values of the modular weights are very restricted. For a classification of
modular weights for all Abelian orbifolds see ref.\cite{IL}.
As a matter of fact, the K\"ahler potentials which appear in the large-$T$
limit of Calabi-Yau compactifications \cite{calabi} and 
4-D fermionic Strings \cite{fermionic}
are quite close to the above one. Thus the results that we will
obtain below will probably be more general than just for orbifold 
compactifications.

Using the particular form (\ref{orbi}) of the K\"ahler potential and
eqs.(\ref{soft},\ref{gaugin}) we obtain
the following results\footnote{This analysis was also carried out for the
particular case of the three diagonal moduli $T_i$
in ref.\cite{japoneses} and \cite{BC}
in order to obtain unification of gauge coupling constants
and to analyze  
FCNC constraints respectively.
Some particular multimoduli examples were also considered in
ref.\cite{FKZ}.} for the scalar masses, gaugino masses and soft trilinear
couplings:
\begin{eqnarray}
&   m_{\alpha }^2 =  \  m_{3/2}^2(1\ +\ 3\cos^2\theta\ {\vec {n_{\alpha }}}.
{\vec {\Theta ^2}}) \ , &
\nonumber\\
&  M = \  \sqrt{3}m_{3/2}\sin\theta e^{-i{\gamma }_S} \ , &
\nonumber\\
&  A_{\alpha \beta \gamma } = \   -\sqrt{3} m_{3/2}\ ( \sin\theta e^{-i{\gamma
}_S}
\ 
+\ \cos\theta \sum _{i=1}^6 e^{-i\gamma _i}    {\Theta }^i {\omega
}^i_{\alpha
\beta \gamma } ) \ , & 
\label{masorbi}
\end{eqnarray}
where we have defined :
\begin{eqnarray}
& {\omega }^i_{\alpha \beta \gamma }\ =\ (1+n^i_{\alpha }+n^i_{\beta
}+n^i_{\gamma
}-
 {Y}^i_{\alpha \beta \gamma }    )\ \ ;\ {Y}^i_{\alpha \beta \gamma } \
= \ {{h^i_{\alpha \beta \gamma }}\over {h_{\alpha \beta \gamma
}}} 2ReT_i \ . &
\label{formu}
\end{eqnarray}
Notice that neither the scalar nor the gaugino masses have any explicit
dependence on $S$ or $T_i$, they only depend on the gravitino mass and
the goldstino angles.
This is one of the advantages of a parametrization in terms of such angles.
Although in the case of the $A$-parameter 
an explicit $T_i$-dependence may appear in
the term proportional to $Y^i_{\alpha \beta \gamma }$, it disappears in 
several interesting cases \cite{BIMS}. With the above information we can now analyze the structure of
soft terms available for Abelian orbifolds. 

{\it 1) Universality of soft terms}

In the dilaton-dominated case ($\cos\theta =0$) the whole
soft terms are universal.
However, in general, they show a lack of universality due to the
modular weight dependence (see eqs.(\ref{masorbi},\ref{formu})).

{\it 2) Soft masses}

In the
multimoduli case, depending on the goldstino direction, tachyons
may appear. For $\cos^2\theta \geq 1/3 $, one has to
be very careful with the goldstino direction if one is interested
in avoiding tachyons. 
Nevertheless, as we will discuss below, having a tachyonic sector is
not necessarily a problem, it may even be an advantage, so one should not
disregard this possibility at this point.

Consider now three particles
$C_{\alpha }$,$C_{\beta }$,$C_{\gamma }$ 
coupling through a Yukawa $h_{\alpha \beta \gamma }$. They may belong
both
to the untwisted (${\bf U}$) sector or to a twisted 
(${\bf T}$) sector, i.e. couplings
of the type ${\bf U}{\bf U}{\bf U}$, 
${\bf U}{\bf T}{\bf T}$,
${\bf T}{\bf T}{\bf T}$. Then, using the above formulae, one
finds \cite{BIMS}
that in general for {\it any choice} of goldstino direction
\begin{equation}
m_{\alpha }^2\ +\ m_{\beta }^2\ +\ m_{\gamma }^2\ \leq \ |M|^2\
=3 m_{3/2}^2\sin^2\theta \ .
\label{rulox}
\end{equation}
Notice that if we insist in having a vanishing gaugino mass, the sum-rule
(\ref{rulox}) forces
the  scalars to be either all massless or at least one of them tachyonic.
Nevertheless we should not forget that tachyons, as we already mentioned
above, are not necessarily a problem, but may just show us an instability.

{\it 3) Gaugino versus scalar masses}

In the multimoduli case 
on average the scalars
are lighter than
gauginos but there may be scalars with mass bigger than gauginos. 
Eq.(\ref{rulox})\ tells us that this can only be true
at
the cost of
having some of the other three scalars with {\it negative} squared mass.
This may have diverse phenomenological
implications depending what is the particle content
of the model, as we now explain in some detail:

{\it 3-a) Gaugino versus scalar masses in standard model 4-D Strings}

Let us suppose we 
insist in
e.g., having tree-level gaugino masses lighter than
the scalar masses.
If we are dealing with a String model with gauge group
$SU(3)_c\times SU(2)_L\times U(1)_Y$$\times G$
this is potentially a disaster.  Some
observable particles, like Higgses, squarks or sleptons would be forced
to acquire large VEV's (of order the String scale). For example, the scalars
associated through the Yukawa coupling $H_2Q_Lu_L^c$, which generates the
mass of the $u$-quark, must fulfil the above sum-rule
(\ref{rulox}). If we
allow e.g. the scalars $H_2$, $Q_L$ to be heavier than gauginos, then
$u_L^c$ will become tachyonic breaking charge and color.
However, tachyons may be helpful if the particular Yukawa coupling
does not involve observable particles. They could break extra gauge symmetries
and generate large masses for extra particles. We recall that standard-like
models in Strings usually have too many extra particles and many extra
U(1) interactions. Although the Fayet-Iliopoulos mechanism helps to cure
the problem \cite{suplemento}, the existence of tachyons is a complementary
solution.

We thus see that, for standard model Strings,
if we want  to avoid charge and colour-breaking minima (or VEV's of
order the String scale for the 
Higgses\footnote{For a possible way-out to this problem, allowing the 
possibility of scalars heavier than gauginos, see ref.\cite{nuevo}.}),
we should grosso modo come back to a situation
with gauginos heavier than scalars.
Thus the low-energy phenomenological predictions of
the multimoduli case are similar to
those of the 
dilaton-dominated scenario (see subsect.2.2):
due to the
sum-rule
the tree-level observable scalars are always ligther than gauginos
\begin{eqnarray}
& m_{\alpha} < M \ . &
\label{masas1}
\end{eqnarray}
Now, at low-energy ($\sim M_Z$), gluino, slepton and (first and second
generation) squark mass relations turn out to be
\begin{eqnarray}
& m_l < m_q \simeq M_g \ , &
\label{masas2}
\end{eqnarray}
where gluinos are slightly heavier than scalars. This 
result is qualitatively similar to the dilaton dominance one, in
spite of the different set of (non-universal) soft scalar masses, because
the low-energy scalar masses are mainly determined by the gaugino loop
contributions. The only exception are the sleptons masses, which do not
feel the important gluino contribution, and therefore can get some
deviation from the result of eq.(\ref{dilaton2}).

As emphasized in \cite{BIM} there is however
a way to get scalars heavier than gauginos,
even in the overall modulus case, if all the observable particles
have  overall modular weight $n_{\alpha}=-1$ and $\sin\theta \rightarrow 0$ (i.e. in the
moduli-dominated limit). Then, at tree-level, $M \rightarrow 0$
and $m_{\alpha}\rightarrow 0$ if the different moduli participate in the
SUSY breaking in almost {\it exactly} the same way, i.e. the overall modulus
situation. Including String loop corrections to $K$ and $f_a$ can yield
scalars heavier than gauginos \cite{BIM}
\begin{eqnarray}
& m_{\alpha} > M_a &
\label{masas3}
\end{eqnarray}
and the low-energy spectrum can be reversed with respect to the above one
(in the case of $\sin\theta$ sufficiently small as to produce
$m_{\alpha} >> M_a$)
\begin{eqnarray}
& M_g < m_l \simeq m_q \ . &
\label{masas4}
\end{eqnarray}
The physical masses of squarks and sleptons are almost
degenerate because the universality of soft scalar masses
at high-energy is not destroyed by the gluino contribution to the
mass renormalization, which is now very small. Notice however that this
possibility of obtaining scalars heavier than gauginos is a sort of
fine-tuning.
In the absence of a more fundamental theory which tells us in what direction
the
goldstino angles point, one would naively say that the most natural possibility
would be to assume that all moduli contribute to SUSY breaking in more or
less (but {\it not} exactly\footnote{For an explicit example
of this, using gaugino condensation, see ref.\cite{Bailin}.}) the same amount.

We just saw how, in the context of standard model Strings,
the results for soft terms are qualitatively similar to the
dilaton dominance ones 
if we want to avoid the breaking of charge and colour conservation.
There is however a loophole in the above analysis.
Up to now we have assumed that the masses of the observable fermions arise
through renormalizable Yukawa couplings. If we give up that assumption
and allow the existence of non-renormalizable Yukawa couplings generating
masses for the observable particles (e.g. $H_2Q_Lu_L^c<\phi...\phi>$), then
new sum-rules would apply to the full set of fields in the
coupling and the above three-particle sum-rules could be violated. In
particular, observable scalars would be allowed to be heavier than gauginos,
possibly at the price of having some tachyon among the (standard model 
singlet) $\phi$ fields. Then qualitative results different from the ones of 
the dilaton dominance case may be obtained

In this respect, it is easy to find explicit examples of orbifold sectors
yielding scalar masses bigger than gaugino masses
even at the tree-level.
 From eq.(\ref{masorbi})
we see that always $m_{\alpha}<m_{3/2}$ and therefore scalars heavier than
gauginos can be obtained if the constraint
\begin{eqnarray}
& \cos^2\theta > 2/3 &
\label{coseno}
\end{eqnarray}
is fulfilled. Let us consider e.g. the case of the
$Z_8$ orbifold with an observable particle in the twisted sector
${\bf T}_{\theta^6}$. The modular weight associated to that sector is
${\vec {n_{\theta^6}}}=(1/4,3/4,0,0)$ and therefore (see eq.(\ref{masorbi}))
\begin{eqnarray}
& m_{\theta^6}^2\ =\ m_{3/2}^2\ \left[1-3\cos^2\theta
\left(\frac{1}{4}\Theta^2_1+\frac{3}{4}\Theta^2_2\right) \right] \ . &
\label{masa}
\end{eqnarray}
For the particular values $\cos^2\theta=5/6$, $\Theta_1=\Theta_2=0$ one
gets $m_{\theta^6}^2=m_{3/2}^2$, $M^2=m_{3/2}^2/2$.

In spite of the new possibilities offered by the multimoduli
extension, one typically finds that, unless very particular choices
for the goldstino angles are chosen,
the masses of scalar and gauginos are still of the same order
and therefore at low-energy eq.(\ref{masas2}) is
typically still valid, the only
difference being that now squarks will be slightly heavier than gluinos.
To reverse the situation (i.e. eq.(\ref{masas4})) we would need
$m_{\alpha}>>M_a$. This can be obtained in the limit
$\sin\theta \rightarrow 0$, i.e. $M \rightarrow 0$.
However, there may be a 
phenomenological problem in this case. Experimental bounds
on gluino mass imply $M>50$ GeV which only can be obtained for a large
$m_{3/2}$ but this would yield a large $m_{\alpha}\sim m_{3/2}$. 
In general one must be careful to avoid
$m_{\alpha}$ bigger than $1$ TeV, spoiling the solution to the
gauge hierarchy problem. 


{\it 3-b ) Gaugino versus scalar masses in GUT 4-D Strings}

What it turned out to be a potential disaster
in the case of standard model Strings may be an interesting
advantage in the case of String-GUTs.
In this case it could well be that
the negative squared mass may
just induce gauge symmetry breaking by forcing a VEV for a particular
scalar (GUT-Higgs field) in the model.
The latter possibility provides us with interesting phenomenological
consequences.
Here the breaking of SUSY would directly induce further gauge symmetry
breaking. 
An explicit example of this situation can be found in ref.\cite{BIMS}.

In summary, the situation concerning gaugino versus scalar masses
is as follows. If any of the physical quark-lepton Yukawas
come from renormalizable terms, the sum rules
leads us to a distribution of soft terms in such a way that
gaugino masses are generically bigger than those of scalars
(otherwise charge and/or colour would be broken). For a possible exception
see footnote 5.
If the physical Yukawas come all from
non-renormalizable terms the constraints coming from
the sum rules may be avoided, possibly allowing standard model singlets
to become tachyonic.
However, even in this case one expects the same order of magnitude
results for scalar and gaugino masses and hence
the most natural (slepton-squark-gluino)
mass relations
{\it at low-energy} will be similar to the ones of
the dilaton-dominated case eq.(\ref{masas2}) as showed in point {\it 3-a}.
Only in the particular limit of very small $\sin\theta$ this situation might be
reversed.

\section{THE $B$ PARAMETER AND THE $\mu$ PROBLEM}

It was pointed out in refs.\cite{GM,CM} that terms in a K\"ahler potential
like the one proportional to $Z_{\alpha \beta }$ in eq.(\ref{kahl})
can naturally induce a $\mu $-term for the $C_{\alpha }$ fields
of order $m_{3/2}$ after SUSY breaking, thus providing a rationale
for the size of $\mu$. 
From eqs.(\ref{kahl},\ref{pot})
and from the fermionic part of the SUGRA lagrangian
one can check that a SUSY mass term 
$\mu_{\alpha \beta} C_{\alpha} C_{\beta}$
and a scalar term 
$B_{\alpha \beta} (C_{\alpha} C_{\beta}) +h.c.$
are induced upon SUSY breaking in the effective low-energy theory
(here the kinetic terms for $C_{\alpha,\beta}$ have 
not still been canonically normalized)
\begin{eqnarray}
& {\mu}_{\alpha \beta} =
m_{3/2} {Z}_{\alpha \beta} - {\overline F}^{\overline{i}} 
\partial_{\overline{i}} {Z}_{\alpha \beta} \ , &
\label{bmu1}
\\
& B_{\alpha \beta} = 
2m_{3/2}^2 {Z}_{\alpha \beta} +    
m_{3/2} F^i \left[ \partial_i Z_{\alpha \beta} - \left(
{\tilde K^{{ \delta} {\overline{\rho}} }}
\partial_i {\tilde K_{{\overline{\rho}}{ \alpha}}} Z_{\delta \beta}
+(\alpha \leftrightarrow \beta)\right)\right]
- m_{3/2} {\overline F}^{\overline{i}} \partial_{\overline{i}} Z_{\alpha \beta}
& \nonumber\\
& - {\overline F}^{\overline{i}} F^j \left[ \partial_j \partial_{\overline{i}}
Z_{\alpha \beta} - \left(
{\tilde K^{{ \delta} {\overline{\rho}} }}
\partial_j {\tilde K_{{\overline{\rho}}{ \alpha}}} \partial_{\overline{i}} 
Z_{\delta \beta}
+(\alpha \leftrightarrow \beta)\right)\right]
\ . &
\label{bmu2}
\end{eqnarray}
Notice that, as in the case of the $A$-terms and the corresponding 
Yukawa couplings (see subsection 2.1),  $B_{\alpha \beta}$
is not necessarily proportional to $\mu_{\alpha \beta}$.

Recently it has been suggested that terms of the type
$Z_{\alpha \beta} C_{\alpha} C_{\beta} +h.c.$
may appear in the K\"ahler potential of some Calabi-Yau type
compactifications \cite{KL}.
It has also been explicitly shown \cite{LLM} 
that they appear in orbifold models.
Let us consider the case in which
e.g., due to gauge invariance, there is only one possible $\mu $-term
(and correspondingly one $B$ term) associated to a pair of matter fields
$C_1$,$C_2$. This is e.g. the case of the MSSM. If we introduce the abbreviations
\begin{equation}
L^Z \equiv  \log Z  \;\; , \;\;
L^{\alpha}  \equiv  \log {\tilde K}_{\alpha }  \;\; , \;\;
X  \equiv  1 - \sqrt{3}  \cos\theta \  e^{i\gamma _i}{\Theta _i}
({\hat K}_{ {\overline i} i})^{-1/2} L_{\overline i}^Z \ ,
\label{xxx}
\end{equation}
using eqs.(\ref{bmu1},\ref{bmu2}) the $\mu$ and $B$ parameters are given by
\begin{eqnarray}
& \mu \ =\ m_{3/2}  ( {\tilde K}_1 {\tilde K}_2 )^{-1/2} Z X \ , &
\label{mmu}
\\
& B\ =\ m_{3/2} X^{-1}
\left[  2 + \sqrt{3} \cos\theta  ({\hat K}_{ {\overline i} i})^{-1/2}
{\Theta_i }
\left( e^{-i\gamma _i}
(  L_i^Z - L^1_i - L^2_i )
-e^{i\gamma _i} L_{\overline i}^Z  \right) \  \right.
& \nonumber\\
& \left. +
\ 3 \cos^2\theta ({\hat K}_{ {\overline i} i})^{-1/2}
{\Theta_i } e^{i\gamma _i} \ \left(
L_{\overline i}^Z ( L^1_j+L^2_j)
- L_{\overline i}^Z L_j^Z - L_{{\overline i} j}^Z\ \right)
({\hat K}_{ {\overline j} j})^{-1/2}
{\Theta _j } e^{-i\gamma _j}      \right] \ , &
\label{bcy}
\end{eqnarray}
where we are assuming that the moduli on
which ${\tilde K}_1(T_i,T_i^*)$, ${\tilde K}_2(T_i,T_i^*)$ and
$Z(T_i,T_i^*)$ depend
have diagonal metric, which is the relevant case we are
going to discuss. The above $\mu$ and $B$ 
(where we have factorized out the $\mu$ term as usual) 
parameters
correspond
now to charged fields which have already been canonically normalized.

If the value of $V_0$ is not assumed to be zero, one just has to replace
$\cos\theta \rightarrow C\cos\theta$ in eqs.(\ref{xxx},\ref{mmu},\ref{bcy}),
where $C$ is given below eq.(\ref{gaugin}).
In addition, the formula for $B$ gets an additional contribution
given by $m_{3/2} X^{-1} 3(C^2-1)$.

As mentioned above, it has recently been shown that
the untwisted sector of orbifolds
with at least one complex-structure field  $U$  possesses the required
structure $Z(T_i,T_i^*)C_1C_2+h.c.$ in their K\"ahler
potentials \cite{LLM}. Specifically, the $Z_N$ orbifolds
based on $Z_4,Z_6$,$Z_8,Z_{12}'$ and the $Z_N\times Z_M$ orbifolds based
on $Z_2\times Z_4$ and $Z_2\times Z_6$ do all have a $U$-type field in (say)
the third complex plane. In addition the $Z_2\times Z_2$ orbifold has $U$
fields in the three complex planes.
In all these models the piece of the K\"ahler potential involving
the moduli and the untwisted matter fields $C_{1,2}$ in the third complex
plane has the form
\begin{eqnarray}
& K(T_i,T_i^*,C_1,C_2)=K'(T_l,T_l^*) 
-\log\left((T_3+T_3^*)(U_3+U_3^*) - (C_1+C_2^*)(C_1^*+C_2)\right)& 
\label{kahlb} \\
& \simeq
 K'(T_l,T_l^*)
 - \log(T_3+T_3^*)  - \log(U_3+U_3^*)\ +
\frac{(C_1+C_2^*)(C_1^*+C_2)}{(T_3+T_3^*)(U_3+U_3^*)} \ .
\label{kahlexp}
\end{eqnarray}
The first term $K'(T_l,T_l^*)$ determines the (not necessarily diagonal)
metric of the moduli $T_l \neq T_3, U_3$ associated to the first and
second complex planes. The last term describes an
$SO(2,n)/SO(2)\times SO(n)$ K\"ahler manifold ($n=4$ if we
focus on just one component of $C_1$ and $C_2$) parametrized by
$T_3, U_3, C_1, C_2$. If the expansion shown in (\ref{kahlexp}) is
performed, on one hand one recovers the well known
factorization $SO(2,2)/SO(2)\times SO(2) \simeq (SU(1,1)/U(1))^2$
for the submanifold spanned by $T_3$ and $U_3$ (which have
therefore diagonal metric to lowest order in the matter fields),
whereas on the other hand one can easily identify the
functions $Z, {\tilde K}_1, {\tilde K}_2$ associated to $C_1$ and $C_2$:
\begin{equation}
Z\ =\ {\tilde K}_1 \ =\ {\tilde K}_2\ =\  {1\over {(T_3+T_3^*)(U_3+U_3^*)}}
\ .
\label{zzz}
\end{equation}
Plugging back these expressions in eqs.(\ref{mmu},\ref{bcy},\ref{xxx})
one can compute $\mu$ and $B$ for this interesting class
of models \cite{BIMS}:
\begin{eqnarray}
& \mu \ =\ m_{3/2}\ \left( 1\ +\ \sqrt{3}\cos\theta
(e^{i \gamma_3} \Theta _3 + e^{i \gamma_6} \Theta _6)\right) \ , & 
\label{muu}
\\
& B\mu=2m_{3/2}^2\ \left( 1\ +\sqrt{3} \cos\theta
 ( \cos\gamma_3 \Theta_3 + \cos\gamma_6 \Theta_6) \ + 
\ 3\cos^2\theta \cos(\gamma_3-\gamma_6) {\Theta _3}{\Theta _6} 
\right) \ . & 
\label{bmu}
\end{eqnarray}
In addition, we recall from eq.(\ref{masorbi}) that the soft masses are
\begin{eqnarray}
& m^2_{C_1}\ =\ m^2_{C_2}\ =\  m_{3/2}^2\ \left( 1\ -\ 3\cos^2\theta
(\Theta_3^2+\Theta _6^2)\right) \ . &
\label{mundos}
\end{eqnarray}
In general, the dimension-two scalar potential for $C_{1,2}$ 
after SUSY breaking has
the form
\begin{eqnarray}
 & V_2(C_1,C_2)\ =\ (m_{C_1}^2+|\mu|^2)|C_1|^2\ 
+ (m_{C_2}^2+|\mu| ^2)|C_2|^2  +(B\mu C_1C_2+h.c.)\ . &
\label{flaty}
\end{eqnarray}
In the specific case under consideration, 
from
eqs.(\ref{muu},\ref{bmu},\ref{mundos}) we find the remarkable result
that the three coefficients in $V_2(C_1,C_2)$ are equal, i.e.
\begin{eqnarray}
& m_{C_1}^2+|\mu|^2 = m_{C_2}^2+|\mu| ^2 = B\mu \ . &
\label{result}
\end{eqnarray}
so that $V_2(C_1,C_2)$ has the simple form
\begin{eqnarray}
& V_2(C_1,C_2)\ =\  B\mu \ (C_1+C_2^*)(C_1^*+C_2) \ . &
\label{potflat}
\end{eqnarray}
Although the common value of the three coefficients in eq.(\ref{result})
depends on the Goldstino direction via the parameters
$\cos\theta$, $\Theta_3$, $\Theta_6$,\ldots (see expression of $B\mu$
in eq.(\ref{bmu})), we stress that the equality itself and the form
of $V_2$ hold {\em independently of the Goldstino direction}.
The only constraint that one may want to impose is that the coefficient
$B\mu$ be non-negative, which would select a region of parameter space.
For instance, if one neglects phases, such
requirement can be written simply as
\begin{eqnarray}
& (1+\sqrt{3} \cos\theta \ \Theta_3) (1+\sqrt{3} \cos\theta \ \Theta_6) \geq 0
\ . &
\end{eqnarray}
We notice in passing that the fields $C_{1,2}$ appear in the
SUSY-breaking scalar potential in the same combination as in
the K\"ahler potential. This particular form may be understood as due to
a symmetry under which $C_{1,2}\rightarrow C_{1,2}+i\delta $ in the K\"ahler
potential which is transmitted to the final form of the scalar potential.

It is well known that, for a potential of the generic form
(\ref{flaty}) (+D-terms), the minimization conditions yield
\begin{eqnarray}
& \sin2\beta  \ =\ { {-2 B\mu} \over {m_{C_1}^2+m_{C_2}^2+2|\mu|^2} } \ . &
\label{sbet}
\end{eqnarray}
In particular, this relation embodies the boundedness requirement:
if the absolute value of the right-hand side becomes bigger than one,
this would indicate that the potential becomes unbounded from below.
As we have seen, in the class of models under consideration
the particular expressions of the mass parameters lead to
the equality (\ref{result}), which in turns implies
$\sin 2\beta= -1$. Thus one finds $\tan\beta=<C_2>/<C_1>=-1$
{\it for any value of $\cos\theta $,$\Theta _3 $,$\Theta _6 $} (and of
the other $\Theta_i$'s of course), i.e. for any Goldstino direction.

As an additional comment, it is worth recalling that in previous
analyses of the above mechanism for generating $\mu$ and $B$
in the String context \cite{KL,BIM,BLM} the value of $\mu$ was left
as a free parameter since one did not have an explicit expression for
the function $Z$. However, if the explicit orbifold formulae for
$Z$ are used, one is able to predict both \cite{BIMS} $\mu$ and $B$ reaching
the above conclusion\footnote{We should add 
that situations are conceivable
where the above result may be evaded, for example if the physical Higgs
doublets are a mixture of the above fields with some other doublets coming
from other sectors (e.g. twisted) of the theory.}.

Now that we have computed explicitly the whole soft terms and the $\mu$
parameter, it would be interesting to analyze the dilaton-dominated scenario
($\cos\theta=0$) because of its predictivity. In particular, from 
eqs.(\ref{dilaton},\ref{muu},\ref{bmu}) we 
obtain\footnote{It is worth noticing here that although the value of $\mu$ is 
compactification dependent even in this dilaton-dominated scenario,
$\mu=m_{3/2}({\tilde K}_1{\tilde K}_2)^{-1/2}Z$ as can be obtained from
eq.(\ref{mmu}), the result 
$\mu=m_{3/2}$ will be obtained in any compactification
scheme with the following property: $({\tilde K}_1{\tilde K}_2)^{-1/2}Z=1$.
Of course, this is the case of
orbifolds, where in particular ${\tilde K}_1={\tilde K}_2=Z$, 
as was shown in eq.(\ref{zzz}).}
\begin{eqnarray}
 & m_{\alpha } = \  m_{3/2} \ , &
\nonumber \\
 & M_a =
 \ \pm \sqrt{3} m_{3/2}   \ , &
\nonumber \\
 & A_{\alpha \beta \gamma } =
 \   - M_a \ , &
\nonumber \\
& B = \  2 m_{3/2} \ , &
\nonumber \\
& \mu = \  m_{3/2} \ , &
\label{dilaton3}
\end{eqnarray}
and therefore the whole SUSY spectrum depends only on one parameter $m_{3/2}$.
If we would know the particular mechanism which breaks SUSY, then we would
be able of computing the superpotential and hence $m_{3/2}=e^K|W|$. 
Although this is not the case,
still this parameter can be fixed from the phenomenological requirement
of correct electroweak breaking $2M_W/g_2^2=<H_1>^2+<H_2>^2$. Thus at the
end of the day we are left with no free parameters. 
Of course, if in the next future the mechanism
which breaks SUSY is known (i.e. $m_{3/2}$ can be explicitly calculated) 
and the above scenario is the correct one, the
value of $m_{3/2}$ should coincide 
with the one obtained from the phenomenological
constraint.
In ref.\cite{nuevo} the consistency of the above boundary conditions with the
appropriate radiative electroweak symmetry breaking is explored. 
Unfortunately, it is 
found that there is no consistency with the measured value of the top-quark
mass, namely the mass obtained in this squeme turns out to be too small.
A possible way-out to this situation is to assume that also the moduli
fields contribute to SUSY breaking since then the soft terms are
modified (see eqs.(\ref{muu},\ref{bmu})) \cite{nuevo}. 
Of course, this amounts to a 
departure of the pure dilaton-dominated scenario.

Finally, let us remark that the previous dramatical conclusion in the
pure dilaton-dominated limit is also obtained in a different
context, namely to avoid low-energy charge and color breaking minima deeper than
the standard vacuum \cite{Amanda}. In fact, on these grounds, the 
dilaton-dominated limit is excluded not only for a $\mu$ term generated
through the K\"ahler potential but for any possible mechanism solving the
$\mu$ problem. The results indicate that the whole free parameter space
($m_{3/2}$, $B$) is excluded after imposing the present experimental data
on the top mass. The inclusion of a non-vanishing cosmological constant
does not improve esentially this situation.

\section*{Acknowledgments}
I thank my collaborators A. Brignole, L.E. Iba\~nez and C. Scheich for an
enjoyable work in this project.

\end{document}